\documentstyle[twocolumn,prb,aps,epsf]{revtex}

\begin{document}

\wideabs{
\title{Coherent dipolar correlations in the ground-state of Kagome frustrated
antiferromagnets.}
\author{Nir Gov}
\address{Department of Physics,\\
University of Illinois at Urbana-Champaign,1110 Green St., Urbana 61801,
U.S.A.}

\maketitle
\tightenlines
\widetext
\advance\leftskip by 57pt
\advance\rightskip by 57pt

\begin{abstract}
We propose a new model for the nature of the low temperature phase
of a geometrically frustrated antiferromgnet (AFM) with a Kagome
lattice, SrCr$_{8-x}$Ga$_{4+x}$O$_{19}$. We propose that the long-range dipolar interaction
between the magnetic Cr$^{3+}$ ions introduces correlations in their
dynamics. The dipolar ground-state has the spins performing correlated zero-point oscillations in a
coherent state with a well defined global phase and a complex
order-parameter (i.e. Off-Diagonal Long Range Order). We calculate
the magnon excitations of such a dipolar array and we find good
agreement with the spin-wave velocities infered from measurements of the specific-heat. Various
experimental properties of these materials are naturally explained
by such a model.
\end{abstract}

\vskip 0.2cm
}

\narrowtext
\tightenlines

\section{Introduction}

The problem of the low temperature magnetic phase of materials with strong
geometric frustration against antiferromagnetic (AFM) order is a long
standing one \cite{chandra}. Especially interesting are systems where there does appear at
finite temperature a low-temperature phase which has some properties of a
spin-glass but also of an ordered spin phase. The most prominent examples are
materials with a lattice containing magnetic ions in planes with Kagome
symmetry \cite{chandra}, such as SrCr$_{8-x}$Ga$_{4+x}$O$_{19}$ (SCGO), on which we shall concentrate in this paper.
Experimental evidence \cite{martinez,martinez94,ramirez92,broholm91,broholm90,aeppli95,lee96prl,lee97} points to a low temperature ($T_{c}\simeq 3$K) phase
which has no static staggered magnetic moment \cite{broholm91}, while at the same time
posesses properties of long-range order \cite{aeppli95} such as a
spin-wave-like spectrum indicated by the specific-heat measurements. There
is additionally a marked difference between Zero-Field-Cooled (ZFC) and
Field-Cooled (FC) magnetic susceptibilities \cite{ramirez94} typical of a
spin-glass (SG) \cite{spinglass}. From
the relaxation rate of polarized muons \cite{amitthes,amit99} it was found that there are locally
fluctuating magnetic spins even at T$\rightarrow $0. Additionally the muon
spin relaxation has a unique Gaussian time dependence \cite{amit99,amit94}.

Taking into account the available experimental data and theoretical
models, we here propose a new model to describe the low-temperature phase of
SCGO. We first note that in addition to the nearest-neighbor exchange
interactions, there is a direct magnetic dipole-dipole interaction between
the Cr$^{3+}$ ions. In the Kagome plane of SCGO this dipolar interaction energy is of
order (taking the Cr$^{3+}$ magnetic moment as \cite{martinez} $3.8\mu _{B}$%
): $E_{dd}=\left( 3.8\mu _{B}\right) ^{2}/{a^{3}}\sim 0.4$K, where the nearest-neighbor distance is \cite{aeppli95} $a=2.93$\AA . This dipolar interaction
is therefore of the order of the transition temperature and is important in determining the properties of the low-temperature phase. This has to be combined with the theoretical analysis \cite{chandra,chalker,broholm91} which points to a high zero-point energy of the Heisenberg Hamiltonian in the Kagome geometry, with no long-range order.

We shall
develope an effective model for the
low-temperature phase of SCGO, where we assume that the spins in the Kagome planes
coherently zero-point oscillate between the different degenerate configurations which minimize the AFM exchange interactions (Fig.1) through
states which minimize the direct magnetic dipolar interaction energy. The
dipolar interaction selects a pair of degenerate configurations which
minimize the overall dipolar energy, and the system then performs coherent zero-point
oscillations between these two equivalent configurations (Fig.2a). We therefore have a quantum
resonance at the frequency determined by the dipolar splitting of the
ground-state. An effective Hamiltonian which describes the coherent dipolar interactions is diagonalized and its spin-wave spectrum agrees with specific heat data.

\section{Dipolar Quantum Resonance}

Above
the transition temperature SCGO has antiferromagnetic correlations with a Curie-Weiss temperature
\cite{ramirez92} of $\theta_{CW} \simeq -500$K, indicating strong
AFM exchange interactions between the magnetic ions.
These strong interactions do not lead to an ordered AFM phase due to
geometric frustration. Extensive numerical calculations of such
two-dimensional spin systems have not yielded any finite temperature phase
transition \cite{chandra,chalker,chubukov}. Classically, this system is characterized by a
highly degenerate and connected (zero-energy modes) ground-state. The quantum
nearest-neighbor Heisenberg model treatment of this system imposes a
constraint of zero total spin in each triangle of the Kagome lattice in the
ground state. Calculations for the purly two dimensional case, show that
quantum fluctuations will select coplanar spin configurations over
non-coplanar ones \cite{chandra}, with the spins tunnelling between the different degenerate configurations \cite{henly}. This does not result though, in a
finite temperature transition to a more ordered phase, due to the geometric frustration.

The ground state of the nearest-neighbor Heisenberg model for the frustrated Kagome AF can be
pictured as spread out over a multi-well spin-space
potential \cite{chandra,chalker}, where the minima of the potential are at the coplanar spin
arrangements (Fig.1). The height of the potential barrier between the minima is of
the order of the AF exchange interaction \cite{henly} $J$ (Fig.1). The
ground state is therefore a superposition of the different coplanar configurations with no long-range order \cite{chandra,chalker,broholm91}, i.e. a "spin-liquid". We
do not wish to attempt a detailed description of this ground-state, which is
a very complicated problem, but rather assume that all the correlations due
to the nearest-neighbor exchange interactions are included in the large (of
order $J$) zero-point energy of the spins in the ground-state. We note that due to the frustration the nearest-nighbor exchange interactions of order $J$ do not freeze the system into one of the minima, but maintain its high zero-point energy (of order $J$), making random zero-point fluctuations between the different states (Fig.1).

Introducing the effect of dipolar interactions, the energies of the different spin configurations over which the ground-state is spread, are now split by the dipolar energy (Fig.1). We assume that the spins in the Kagome planes
make coherent zero-point oscillations between the different degenerate "J-minimizing" configurations (Fig.1) through
states which minimize the direct magnetic dipolar interaction energy. The
dipolar interaction selects a coherent state out of the ground-state
superposition of spins directions (Fig.2b). In this coherent state the
individual spins zero-point oscillate between the many coplanar (J-minimizing) states (Fig.1) through intermediate states that
minimize the dipolar interaction energy (Fig.2a). The relatively weak
dipolar interaction splits the Heisenberg ground-state energy and creates a
situation of macroscopic quantum resonance. We therefore have a collective `double well' state, in which many nearly
degenerate spin configurations reach resonance due to the coupling through
the dipolar interaction.
In this quantum resonance the system makes zero-point correlated oscillations between J-minimizing
configurations through the barrier configurations (Fig.1) that minimize the dipolar interaction energy (Fig.2). The quantum
resonance is at a frequency determined by the dipolar splitting of the
ground-state.

The large difference in energy scale between the exchange ($\sim 100$K) and dipolar
interactions ($\sim 1$K) allows us to limit our treatment to the dipolar interactions
alone. Assuming therefore that the coherent zero-point oscillation of the spins is
controlled by the relatively weak dipolar interactions, we can
consider only these interactions when describing the low-lying excitations
of the low-temperature phase. This allows us to proceed by describing the spin-waves as the
modulations in the dipolar interactions around the configuration that
minimizes these interactions alone.

The specific spin configuration which we find that minimizes \cite{minimum} the dipolar
interaction is shown in Fig.2a. The additional constraint of zero total magnetic moment in each unit cell, is fullfilled by alternating between the up/down spins in each of the interpenetrating Kagome planes of the SCGO crystal \cite{broholm90}.
This is a not a minimum configuration with respect to  
the exchange interactions, since the total spin on each triangle is not zero, but the system explores these states due to its high zero-point energy of order $J$ (Fig.1). The overall dipolar interaction energy is given by
\begin{equation}
E_{dd}=\sum_{i\neq j}\frac{{\bf \mu }_{i}\cdot {\bf \mu }_{j}-3({\bf \mu }%
_{i}\cdot \widehat{{\bf r}}_{ij})({\bf \mu }_{j}\cdot \widehat{{\bf r}}_{ij})%
}{\left| {\bf r}_{ij}\right| ^{3}}  \label{edipole}
\end{equation}
where for Cr$^{3+}$ ions in SCGO, $\left| {\bf \mu }\right| =3.8\mu _{B}$.
This energy is $E_{dd}\simeq -1.6$K for the spins marked by filled circles (A) and $E_{dd}\simeq0.4$K for
spins marked by empty circles (B) in Fig.2a, giving an overall energy reduction in this
arrangement, since there are two B spins for every A spin. The local flip ($%
\pi $ phase shift) of an A spin out of the above arrangement costs an
energy $E_{0}=2\left| E_{dd}\right| \simeq 3.2$K, and can be treated as a local
excitation. This is just the resonance energy split shown in Fig.1,
describing the zero-point oscillation of the spins between the two equivalent up-down
configurations of Fig.2. The experimental data indicates almost no frozen (static) magnetic moment \cite
{broholm91,amit94} at low temperatues, justifying taking the full magnetic
moment of the Cr$^{3+}$ ion as taking part in the coherent zero-point oscillations. We point out that in the two-dimensional Kagome planes there is
a global axis along which the spins naturally resonate, which is the normal
to the planes. This is in contrast to a three-dimensional pyrochlore network
material, where there is no such global axis.

The above model therefore
accounts for the dynamic nature of the spins in the low-temperature phase,
while having a long-range phase order. 
Experimental evidence for this proposed arrangement of the zero-point oscillating spins may be indicated by the
diffuse elastic-Bragg scattering peak detected in neutron scattering
\cite{broholm91}. The measurement indicates a lack of a
well-defined long-range spatial order. In our model the spins zero-point oscillate in
phase over the entire lattice, but different crystals in the powder sample
have different phases, so that the overall interference is randomized. Furthermore, the J-minimizing states through which the spins zero-point oscillate are random between different Kagome planes, i.e. have no long-range spin order. The
peak position at $\sim $1.4\AA $^{-1}$ corresponds to a periodicity of $\sim
$4.5\AA , which is the size of approximately two triangles in the Kagome
plane, and agrees with the periodicity of the dipolar arrangement we
propose in Fig.2a. We wish to stress that the dynamic nature of this coherent-state involves a gauge symmetry breaking, in the form of a defined relative phase of the zero-point oscillating dipoles (see Eq.(\ref{psi0})).
The time-independent ground-state is described by averaging over all possible global phases, after establishing the relative phase relation. This is similar to the case of superconductivity and superfluidity \cite{forster}.

An additional advantage of our model is that the energy scale of the
low-temperature phase is determined by the long-range (dipolar) interactions.
This means that the absence of any measured critical behavior at the Kagome
percolation concentration \cite{amit99} $p_{percol}=0.6527$ of the Cr atoms,
is naturally explained. A phase transition driven by the nearest-neighbor
exchange interactions would have been sensitive to the percolation
transition. The linear dependence of the transition temperature \cite
{martinez} on the Cr concentration $p$ also follows naturally from the
summation in (\ref{edipole}). It was previously noted \cite{huber} that the
independence of the qualitative properties of the low-temperature phase on the dilution $p$ may indicate
the occurence of long-range interactions.
From experiments \cite{martinez} it is found that the zero transition temperature is shifted to a critical
dilution $p_{c}\simeq 0.2$. This non-zero dilution may arise due to some
additional weak interactions or impurities outside the Kagome planes, which
dominate over the dipolar interactions when the latter become too weak, and therefore
destroy the coherent state below $p_{c}$.

The coherent order of the zero-point oscillating spins can be destroyed if an
additional static local magnetic moment induces a preferred static
orientation for the Kagome spins. Such experiments \cite{martinez94} show
that magnetic ions (Fe) in the layers between the Kagome planes, turn the
SCGO into a normal SG material. These static (but random) magnetic ions
destroy the quantum resonance of the Kagome spins and force them into a
static equilibrium orientation. The system has a new SG transition
temperature $T_{SG}\sim 25$K, i.e. an order of magnitude larger than the
dipole-induced transition temperature $T_c$. This indicates that
the SG transition is controlled by the strong exchange interactions between
the Cr-Cr and Cr-Fe electrons, which induce a static freezing of the spins
in random orientations.

\section{Spin-waves}

The collective excitations of the coherent zero-point oscillating spins are spatial
modulations of the relative phases of the spins with respect to the ground-state configuration of Fig.2a. The spins in the sparse rows (filled circles in Fig.2a) are in a
minimum of the dipolar energy so that they feel a restoring force and
support spin-wave excitations. The effective Hamiltonian describing the
interacting local spins, taking into account only the dipolar interaction, is 
\cite{bcc,anderson}
\begin{eqnarray}
{H_{loc}} &=&{\sum_{k}}(E_{0}+X(k))\left( {{b_{k}}^{\dagger }}{b_{k}}+{\frac{%
1}{2}}\right)   \nonumber \\
&&\ \ +{\sum_{k}}X(k)\left( {{b_{k}}^{\dagger }}{b_{-k}^{\dagger }}%
+b_{k}b_{-k}\right)   \label{hloc}
\end{eqnarray}
where ${{b_{k}}^{\dagger },}{b_{k}}$ are Bose creation/anihilation operators
of a local spin-flip with respect to the configuration of Fig.2a. These local
spin-flips can be treated as bosons using the standard Holstein-Primakoff
procedure \cite{anderson}. $E_{0}$ is the bare energy of a local spin flip and $X(k)$ is the
dipolar interaction matrix element modulated along some direction ${\bf k}$
in the Kagome plane, given by \cite{heller}
\begin{eqnarray}
X\left( {\bf k}\right)  &=&-\left| {\bf \mu }\right| ^{2}\sum_{i\neq
0}\left[ \frac{3\cos ^{2}\left( {\bf \mu }\cdot \left( {\bf r}_{0}-{\bf r}%
_{i}\right) \right) -1}{\left| {\bf r}_{0}-{\bf r}_{i}\right| ^{3}}\right]
\nonumber \\
&&\times \exp \left[ 2\pi i{\bf k}\cdot \left( {\bf r}_{0}-{\bf r}_{i}\right)
\right]   \label{xk}
\end{eqnarray}
where we assume that the ground-state is given by the configuration of
Fig.2, so that all the magnetic moments ${\bf \mu }$ are normal to the Kagome planes.

At $k=0$ the interaction matrix $X(k)$ is just the dipolar energy (\ref
{edipole}). The Hamiltonian ${H_{loc}}$ (\ref
{hloc}), which describes the effective interaction between localized modes,
can be diagonalized using the Bogoliubov transformation ${\beta _{k}}%
=u(k)b_{k}+v(k)b{^{\dagger }}_{-k}$. The two functions $u(k)$ and $v(k)$ are
given by
\begin{equation}
{u^{2}}(k)={\frac{1}{2}}\left( \frac{E_{0}{+X(k)}}{{E(k)}}+1\right) ,{v^{2}}%
(k)={\frac{1}{2}}\left( \frac{E_{0}{+X(k)}}{{E(k)}}-1\right)  \label{uv}
\end{equation}
The coherent ground state is given by \cite{huang}
\begin{equation}
\left| \Psi _{0}\right\rangle =\prod_{k}\exp \left( \frac{v_{k}}{u_{k}}{{%
b_{k}}^{\dagger }}{b_{-k}^{\dagger }}\right) \left| vac\right\rangle
\label{psi0}
\end{equation}
and the energy spectrum is
\begin{equation}
E(k)=\sqrt{E_{0}\left( E_{0}+2X(k)\right) }  \label{ek}
\end{equation}
It is clear from (\ref{ek}) and the definition of $E_{0}$ that
the spectrum of the excitations is gapless, since we have $-2X(0)\equiv
E_{0}\simeq 3.2$K, i.e. the bare local-mode energy is a local spin flip \cite
{bcc}.

The function $X(k)$ can be calculated in any direction of the lattice with the corresponding energy spectra (\ref{ek}) (Fig.3). Since
spin-waves are transverse and our ground-state configuration has the spins
normal to the Kagome planes (Fig.2), we have spin-waves only in the Kagome
planes. We find that in the limit $k\rightarrow 0$ the energy spectrum is
linear with a velocity in the range 60-80 m/sec (Fig.3). The specific heat
measurements provide an estimate of the excitation spectrum in this linear limit. The
spin wave velocity $C$ at $k\rightarrow 0$ is related to the specific-heat by
\cite{ramirez92}: $C_{v}=2.3\left( k_{B}^{3}/h^{2}C^{2}\right) T^{2}$. This velocity is found to be linear \cite{ramirez92} in the dilution $p$ and is $\sim 100$ m/sec ($p=0.89$).
This velocity is much lower than that calculated using the
strong nearest-neighbor exchange interactions \cite{chubukov,huber}.
Current neutron scattering data \cite{aeppli95} is not accurate enough at low energies to resolve the detailed structure of the spin-wave spectrum. Still in Ref.[6] there is some structure in the inelastic neutron scattering at energies $\sim 2-5$K, which may indicate the spin-wave spectrum of Fig.3.

To compare with our calculation we must divide our calculated specific heat
by 3 since only a third of the spins reside in the sparse rows (Fig.2), for
which the spin waves are described by (\ref{ek}). To agree with the experimental
data we would therefore need a velocity of $\sim 60$ m/sec, which is indeed in the
range of velocities we calculated. The linear dependence of this velocity on
the dilution $p$ is again a trivial consequence of (\ref{edipole}), since
there is a single energy scale in our model.

Using the analogy with a usual AFM \cite{kittel} (see next section), a mean-field description of the thermal reduction of the staggered magnetization gives a transition temperature: $T_{c}\sim 3$K for the sample with $E_{0}\simeq 2.2$K ($p=0.89$). This is in rough agreement with the measured transition temperature \cite{ramirez92} of $3.5\pm 0.1$K.

\section{Spin Correlations and Experimental Probes}

The configuration of the relative phases of the zero-point scillating dipoles in the ground-state of Fig.2a resembles an AFM.
We shall now make the analogy between the coherent ground-state (Fig.2) and an AFM more precise.
The operators ${{b_{k}}^{\dagger },}{b_{k}}$, of a spin-flip are with respect to the ground-state configuration of Fig.2a, and therefore correspond to the staggered magnetization operators of an AFM (in the small $k$ limit). Their correlation function:
$S_{{b_{k}},{b_{-k}}}(k)=(v(k)-u(k))^2=E_{0}/E(k)$, has the $1/k$ divergence expected for the long-range order of the staggered magnetization in an AFM \cite{forster}. 
The static structure-factor of the spins, as measured by neutron scattering \cite{lee96prl}, shows approximately a linear behavior: $S(k)\sim k$ in the $k\rightarrow 0$ limit. This is typical of the response function of the transverse magnetization in an AFM \cite{forster}. 
The total transverse magnetization behaves as the density of a normal liquid, and in the limit $k\rightarrow 0$ has a correlation function: $S_{M_{\perp}M_{\perp}}(k\rightarrow 0)\rightarrow \hbar ck/E_{0}$ (where we wrote the dimensionless structure-factor using the perpendicular susceptability $\chi_{\perp}\sim 1/E_{0}$).
The measured dynamic response functions are therefore in agreement with this model, in which the ground-state has some features of a standard AFM. Unlike a static AFM the staggered magnetization operators ${{b_{k}}^{\dagger },}{b_{k}}$ are with respect to the zero-point coherent oscillations, driven by the dipolar interactions of energy $E_{0}$. Their spatial correlation functions, though, are similar.

Another puzzling phenomenon of the SCGO is the marked difference between
Zero-Field-Cooled (ZFC) and Field-Cooled (FC) static magnetic susceptibility \cite
{ramirez94,ramirez92}. The measured cusp in the ZFC magnetic susceptibility defines the transition
temperature. These experimental results resemble the difference between the
longitudinal and perpendicular susceptibilities of a usual AFM \cite{kittel}%
, and also of a normal SG material \cite{spinglass}. We shall now give a qualitative
description of this behavior in SCGO, as follows from our model of the
low-temperature phase.

In the FC case we have a magnetic field (taken to be along the $z$%
-direction) which breaks the quantum resonance for the zero-point oscillating spin
configurations in planes perpendicular to the field (Fig.4b). Since the
experiment is done using a powder, there are crystals with all
possible orientations of the Kagome planes. The coherent phase (i.e., the
quantum resonance) therefore develops only in planes which are parallel to
the external field, where the spins have the zero-point oscillating AFM-like
order of Fig.2a, perpendicular to the applied field. The planes
without the coherent order give zero average instanteneous contribution to
the internal magnetic field in the $z$-direction (for relatively weak fields $\mu H << J$). We therefore
expect to find a magnetic response which is similar to that of a normal AFM
in a perpendicular external magnetic field, which is non-zero and almost constant with
temperature: $\chi_{FC}=\chi_{\perp}\sim 1/E_{0} \neq 0$. The smallness of the dipolar interactions ($E_{0}$) account for the relatively large low-temperature FC susceptibility.

In the ZFC state the coherent order is established in all the crystals of
the powder (Fig.4a). For an applied field which is weak compared with the
internal magnetic fields, spins parallel to the external field do not
respond \cite{kittel} ($\chi _{\parallel }\rightarrow 0$ for T$\rightarrow 0$). The internal
fields in the $z$-direction due to these spins
completely mask the external field so that crystals with spins perpendicular
to the external field do not respond either, i.e., $\chi_{ZFC}(T=0)= \chi_{\parallel }(T=0)=0$.
This explains the vanishing
susceptability of the ZFC in small applied fields as T$\rightarrow 0$. The
magnitude of the internal magnetic fields can be estimated from (\ref
{edipole}) as: $H_{0}\simeq E_{0}/3.8\mu _{B}\simeq 12$kG. Only for external fields
approaching the size of these internal fields does the T=0 ZFC response
approaches the FC response \cite{amitthes,amit99}.
 
Muon Spin-Relaxation ($\mu $SR) experiments \cite{amitthes,amit99} probe the spatial and temporal correlations of the spins. These experiments show that the
low-temperature relaxation rate of polarized muons is finite and
temperature-independent in the low-temperature phase, which means that the
magnetic spins are fluctuating even at T$\rightarrow $0. The coherent state
we propose has long-range spatial correlations of the phase of the
zero-point oscillating spins, so they do remain dynamic even at T=0.

Indeed, (\ref{psi0}) describes a coherent state with Off-Diagonal Long-Range
Order (ODLRO) of the zero-point oscillating spins. This is a system with broken global gauge
symmetry, in the form of a global phase of the zero-point oscillating spins (a complex
order parameter). These long-range spatial and temporal
correlations appear in the time dependence of the muon spin relaxation which
changes from exponential above the transition temperature to Gaussian \cite{amit99} as T$%
\rightarrow $0. Gaussian decay of the polarization arises in
cases of long-range (time-independent) temporal correlations between the spins\cite{budnick}, as occurs in the coherent phase we propose. The coherent zero-point oscillations of the spins will produce a zero average static magnetic moment, so there will therefore be no oscillating signal in the muon polarization decay, which is typical of static magnetic order \cite{budnick}. The muon spins that will be excited to oscillate in phase with the coherent zero-point oscillations of the Kagome spins (i.e. frequecy $E_{0}/\hbar$), will see a constant magnetic field, with a resulting Gaussian time decay of the muon polarization \cite{nmr}.
The rate at which muons will follow the coherently oscillating Kagome spins is given by second-order time dependent perturbation theory \cite{cohen}. It is equal to the probability per unit time of exciting the muon by the periodic zero-point oscillations of the lattice spins
\begin{equation}
P(t) \propto \frac{\left|W\right| ^2}{\hbar ^2 \omega}
\end{equation}
where the the matrix element coupling the muons to the local magnetic fields is $\left|W\right| \simeq \gamma _{\mu }\Delta B_{\mu }$ where $\Delta B_{\mu }$ is the rms deviation of the
internal magnetic field at the muon site and $\gamma_{\mu }$ is the muon magnetic moment. Since this rms field is due to the
coherent zero-point oscillations of the Kagome spins, it is of order $H_{0}$. The frequency $\omega=E_{0}/\hbar$ is that of the zero-point oscillations of the lattice spins, so the resulting rate of muon depolarization is given by
\begin{equation}
\lambda=\frac{\vert \gamma _{\mu }\Delta B_{\mu } \vert^2}{E_{0}} \simeq 10 \rm {MHz} \label{lambda}
\end{equation}
in excellent agreement with experimental results \cite{amit99}.
This rate of muon depolarization should apply as long as the probability is much less than 1, i.e. for $t  \ll 1/\lambda \simeq 0.1 \mu$sec. Indeed at longer times there is a deviation from the Gaussian depolarization curve \cite{amit99}.
Since the coherent zero-point magnetic fields are linear with the dipolar energy $H_{0}\propto E_{0}$, this decay rate is linearly
proportional to the dilution $p$.
The linear dependence of $\lambda$ on the dilution $p$ is roughly supported by the experimental results \cite{amit99}, if we note that the zero of the coherent oscillations is shifted to \cite{martinez} a critical
dilution $p_{c}\simeq 0.2$, as does the transition temperature.
Above the
transition temperature the spin fluctuations are uncorrelated, resulting in an
exponential decay of the muon polarization.

\section{Conclusion}

We conclude that a model of correlated and coherent zero-point oscillations of the Kagome spins, driven by the magnetic
dipolar interactions, describes the essential features of the low-temperature phase of the geometrically
frustrated AFM with Kagome lattice, namely SCGO. This model describes a magnetic phase which has an ODLRO and a complex order parameter (\ref{psi0}). If the disordered high-temperature phase is called a spin-liquid \cite{chandra}, then the coherent low-temperature phase described in this work is a "spin-superfluid". The number occupation of the coherent spin-flips diverges: $\left\langle n_{k}\right\rangle =2v(k)^2 \sim 1/k$, signaling condensation in the $k=0$ state in a Bose liquid \cite{gavoret}.

Due to the complex order-parameter, this phase can therefore support linear defects ("spin-vortices") in this order parameter, with quantized "spin-currents".
The coherent state we have proposed may also have relevance to the studies of macroscopic quantum coherence in molecular magnets and other materials \cite{molmag}.

{\bf Acknowledgement} Part of this work was done in the
Technion-Israel Institute of Technology, 
Haifa, Israel. I thank Gordon Baym, Amit Keren and Efrat Shimshoni for usefull discussions and suggestions. I also thank Amit Keren for access to his recent experimental data.
This work was supported by
the Fulbright Foreign Scholarship grant, the Center for Advanced Studies and
NSF grant no. PHY-98-00978.

\begin{figure}[tbp] 
\input epsf \centerline{\ \epsfysize 5.5cm \epsfbox{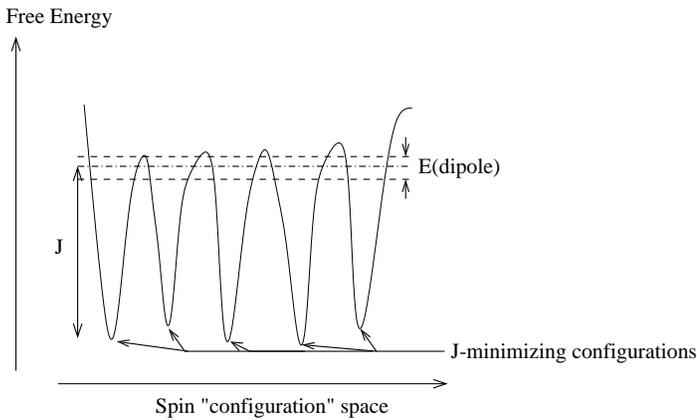}}
\vskip 3mm
\caption{Schematic description of the free energy surface of the spin configuration on the Kagome lattice [1]. The ground-state of the exchange interactions ($J$, dash-dot line) is not localized in any particular minima due to the geometric frustration. The dipole-induced splitting of the ground-state (Eq.(1)) is indicated by the pair of dashed lines.} 
\end{figure}

\begin{figure}[tbp] 
\input epsf \centerline{\ \epsfysize 5cm \epsfbox{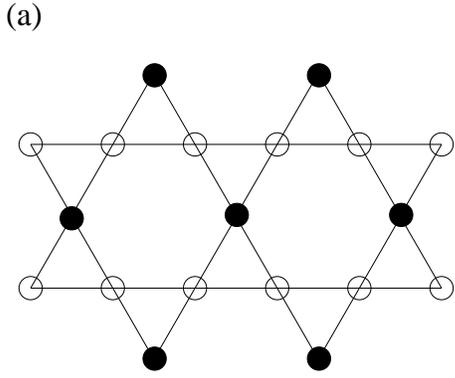}}
\vskip 2mm
\input epsf \centerline{\ \epsfysize 3.5cm \epsfbox{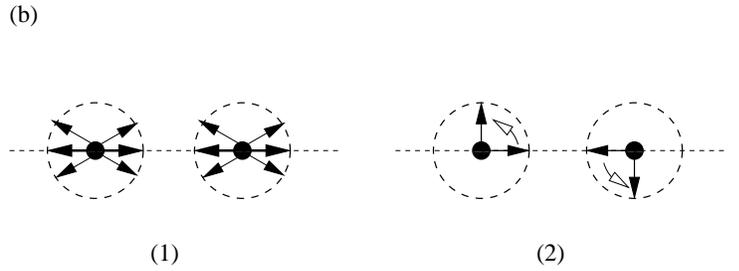}}
\vskip 3mm
\caption{(a) The minimum configuration for the spins with respect to the dipolar interactions in the Kagome planes (the plane of the figure). The spins are normal to the Kagome planes, with up/down spins corresponding to filled/empty circles respectively. The system makes zero-point coherent oscillations between this and its spin-reversed configuration.
(b) Schematic illustration of the spins in the non-coherent superposition ground-state ("spin liquid") due to the exchange interactions alone (1), and making zero-point oscillations in synchrony between the up/down configurations of (a) in the dipole-induced coherent state (2). The straight dashed lines represent the Kagome planes, viewed sideways.} 
\end{figure}

\begin{figure}[tbp] 
\input epsf \centerline{\ \epsfysize 8.5cm \epsfbox{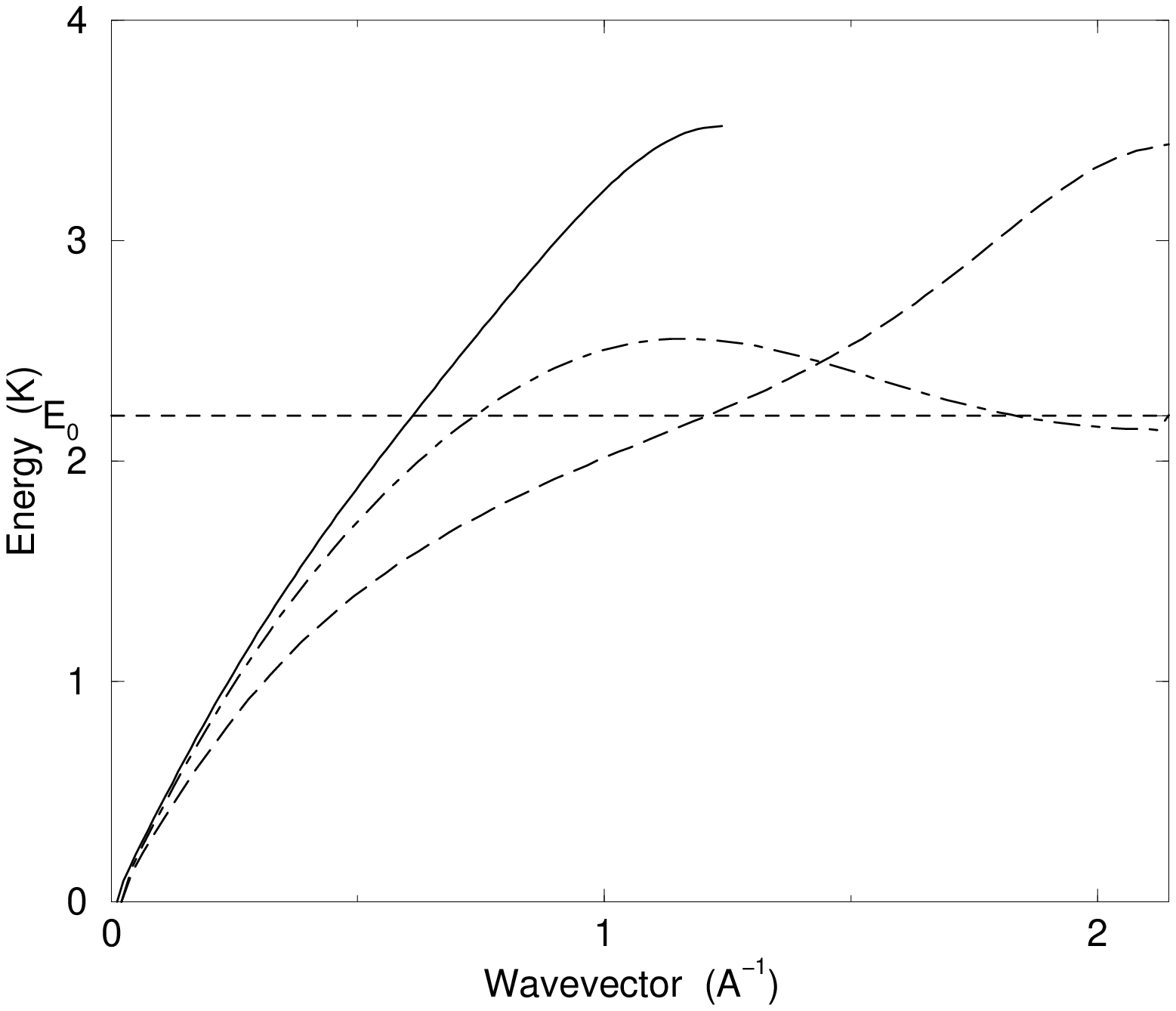}}
\vskip 1mm
\input epsf \centerline{\ \epsfysize 3.5cm \epsfbox{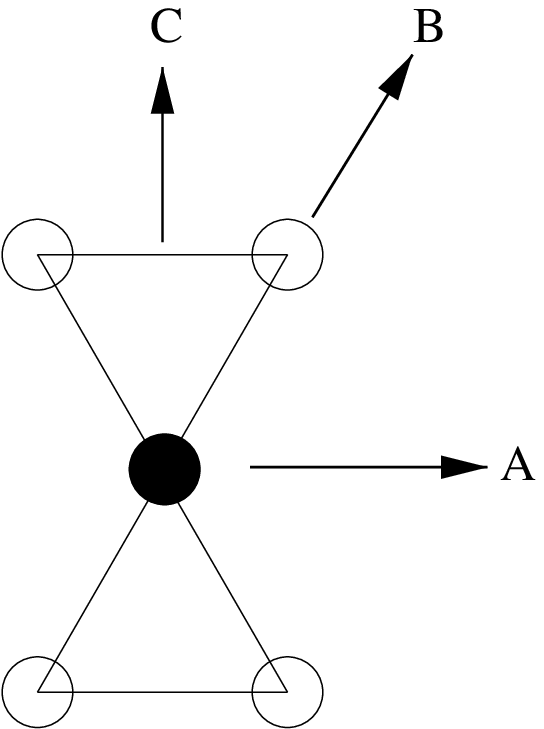}}
\vskip 3mm
\caption{The calculated spectrum of the spin waves in the Kagome planes, using Eq.(6), for dilution $p=0.89$, where $E_{0}\simeq 2.2$K. The different directions along which the spectrums have been calculated are shown below: A- dash line, B- dash-dot line, C- solid line.} 
\end{figure}

\newpage

\begin{figure}[tbp] 
\input epsf \centerline{\ \epsfysize 6.5cm \epsfbox{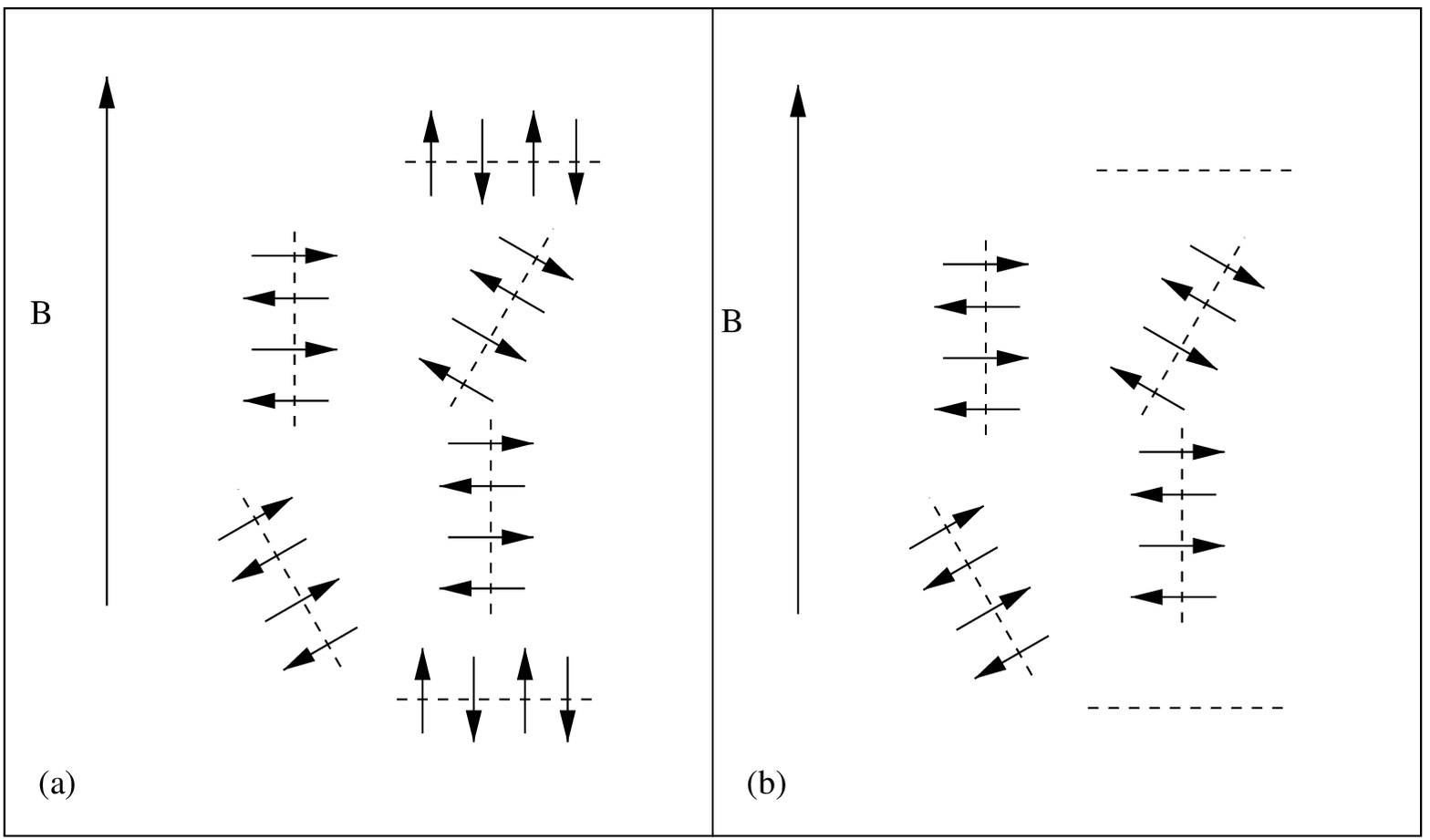}}
\vskip 3mm
\caption{Schematic picture of the Kagome planes (broken lines) in the different crystal
grains, with the coherent spins
indicated. (a) The ZFC case where all the Kagome
planes are in a coherent state.
(b) The FC case where planes perpendiclar to the field
have no long-range coherent order (dashed planes with no arrows). $B$ is the
external magnetic field, with the direction indicated by the long arrow.} 
\end{figure}


\begin{references}
\bibitem{minimum} The true minimum of the dipolar interactions can at most be different from that calculated using the configuration of Fig.2 by a factor of $\sim$2. This upper bound can be estimated by calculating the (known, unfrustrated) dipolar energy of a simple-cubic lattice with the same density and layer structure.
The frustration of the Kagome lattice should greatly lower this upper bound.
\end{references}
\end{document}